\documentclass{ifacconf}

\usepackage{graphicx}      
\usepackage{natbib}        
\usepackage{amsmath,amssymb,amscd}
\usepackage{bm}
\usepackage{float}
\usepackage{color,url}
\usepackage{stfloats}
\usepackage{subfigure}
\usepackage{braket}

\newtheorem{lemma}{Lemma}
\newtheorem{theorem}{Theorem}
\begin{document}
\begin{frontmatter}

\title{Robustness analysis in static and dynamic quantum state tomography}


\thanks[footnoteinfo]{This research was supported by the Australian Research Council Future Fellowship Funding Scheme under Project FT220100656. \emph{(Corresponding author: Shuixin Xiao)}.}

\author[1]{Alan Chen},    
\author[1,2]{Shuixin Xiao}, 
\author[1,3]{Hailan Ma},
\author[4]{Daoyi Dong}


\address[1]{School of Engineering, Australian National University, Canberra ACT 2601, Australia (e-mail: zihanchen0408@gmail.com, shuixin.xiao@anu.edu.au)} 
\address[2]{Department of Electrical and Electronic Engineering, University of Melbourne, Parkville VIC 3010, Australia}   
\address[3]{Nanyang Quantum Hub, Nanyang Technological University,  637371, Singapore (e-mail:hailanma0413@gmail.com)}
\address[4]{Australian Artificial Intelligence Institute, Faculty of Engineering and Information Technology, University of Technology Sydney, NSW 2007, Australia (e-mail: daoyidong@gmail.com)}

\begin{abstract}                

Quantum state tomography is a core task in quantum system identification. Real experimental conditions often deviate from nominal designs, introducing errors in both the measurement devices and the Hamiltonian governing the system’s dynamics. In this paper, we investigate the robustness of quantum state tomography against such perturbations in both static and dynamic settings using linear regression estimation. We derive explicit bounds that quantify how bounded errors in the measurement devices and the Hamiltonian affect the mean squared error (MSE) upper bound in each scenario. Numerical simulations for qubit systems illustrate how these bounds scale with resources. 
\end{abstract}

\begin{keyword}
Quantum system identification, quantum state tomography, robustness analysis
\end{keyword}

\end{frontmatter}

\section{Introduction}
Over the past few decades, various quantum technologies have made remarkable progress, including quantum computing~\citep{Nielsen_Chuang_2010}, quantum communication \citep{Gisin_2007}, quantum sensing~\citep{Degen_2017}, and quantum control~\citep{Dong_Petersen_2010, Dong_Petersen_2022, Dong_Petersen_2023}. In these applications, a fundamental task is to design effective estimation and identification algorithms to obtain complete information about the quantum system of interest. This process is called quantum tomography~\citep{Nielsen_Chuang_2010} or quantum system identification~\citep{Burgarth_Yuasa_2012, ma1}.  In this work, we focus on quantum state tomography (QST), where the goal is to estimate parameters to reconstruct an unknown quantum state using measurement data.

In quantum science and technology, given known positive operator-valued measures (POVMs), QST is the process of inferring an unknown quantum state from such measurement data~\citep{Paris_2004}. Common algorithms for QST include Maximum Likelihood Estimation~\citep{Hradil_1997}, Bayesian Mean Estimation~\citep{Blume-Kohout_2010} and Linear Regression Estimation (LRE)~\citep{Qi_2013}. Since quantum states often have low rank, \cite{Gross_2010} proposed QST via compressed sensing, and \cite{Mu_2020} leverages regularization to use fewer measurements. Collective methods have been studied in \cite{11158864,zhou2023experimental,xiao2025gen}.
Among these algorithms, LRE stands out as a scalable method whose asymptotic mean squared error (MSE) upper bound is independent of the true quantum state being reconstructed. 

In addition to the standard LRE setting based on static measurement outcomes, recent studies have shown that the time evolution of observables can also be used to reconstruct an unknown quantum state~\citep{Kech_2016, Xiao_2024, Rall_2025}. Related work formulates the problem from a control-theoretic observability perspective~\citep{Kech_2016, Rall_2025, Peruzzo_2024, Peruzzo_2025}. In this formulation, the measurement data consist of time-dependent expectation values under a known Hamiltonian, and the reconstruction problem is cast in a state-space model that connects identifiability of the initial state to the observability properties of a linear dynamical system~\citep{Xiao_2024}.

In theoretical formulations of QST, the measurement operators, often represented as a set of POVMs, are assumed to be perfectly known and accurately implemented. However, in realistic experimental conditions, these operators may suffer from sources of errors and imperfections, such as misalignment or miscalibration~\citep{Lundeen_2009}. Such deviations introduce perturbations to the intended measurement operators, leading to systematic errors in the recorded statistics. Since these perturbations directly alter the measurement operators used in reconstruction algorithms, they can in turn affect the accuracy of the estimated quantum state and its associated MSE.

Building on these ideas, this work investigates the robustness of quantum systems against perturbations in both static and dynamic settings within the LRE framework. In the static case, we analyze perturbations to the measurement operators; in the dynamic case, we study perturbations to the observable and the system Hamiltonian jointly. We provide explicit conservative bounds that characterize the impact of bounded perturbations on the MSE upper bound in each setting. These bounds describe how the MSE upper bound grows with the error magnitudes, the system dimension, the sampling number and schedule (in the dynamic setting), the number of measurement configurations, the number of available copies, and the conditioning of the design or observability matrix through its smallest eigenvalue.

The organization of this paper is as follows. Section \ref{sec2} introduces preliminary knowledge on QST using LRE and observable evolutions. Section \ref{sec3} presents the robust analysis on how the proposed perturbations propagate through the system. In Section \ref{sec4}, we use numerical simulations to validate and illustrate the various analytical error bounds, and Section \ref{sec5} concludes this paper.

\section{Preliminaries}\label{sec2}

This section provides a brief introduction about QST via the LRE and observable evolutions. For a quantum system with $d$-dimensional Hilbert space, we use $\{E_i\}_{i=1}^{d^2}$ as a complete set of operators that form an orthonormal basis. Specifically, it satisfies $\operatorname{Tr}\left( E_i^\dagger E_j \right)=\delta_{ij}$, where $\dagger$ denotes the Hermitian adjoint and $\delta_{ij}$ is the Kronecker function. Let $E_i=E_i^\dagger$ (Hermitian) and $E_{d^2}=\frac{I}{\sqrt{d}}$ where the other bases are traceless.

In the system, quantum states can be represented using a density operator $\rho$ that is Hermitian ($\rho=\rho^\dagger$), positive semidefinite ($\rho\geq0$) and has unit trace ($\operatorname{Tr}(\rho)=1$). 
To gain information about an unknown state, measurement devices are required, known as detectors. These detectors can be characterized by a set of operators $\{P_i\}^n_{i=1}$, which are the aforementioned POVMs. Importantly, together the POVM set satisfies the completeness constraint $\sum^n_{i=1}P_i=I$. 
When the measurement operators $\{P_i\}$ are applied to the quantum state $\rho$, the ideal probability of obtaining the $i$-th result is given by Born's rule,
\begin{equation}
    \label{born}
    p_i=\operatorname{Tr}(P_i\rho).
\end{equation}
If $N$ copies of the unknown state $\rho$ are prepared and the $i$-th outcome occurs $N_i$ times, then the estimate of $p_i$ is given by $\hat{p}_{i}=N_i/N$ and the corresponding measurement error is $e_i=\hat{p}_i-p_i$.
By the central limit theorem~\citep{Qi_2013, Chow_2012}, the error converges to a normal distribution satisfying $e_i \sim \mathcal{N}\left(0,\frac{p_i-p_i^2}{N}\right)$.

\subsection{Static tomography via linear regression estimation}\label{qst_lre}

Using the chosen basis $\{E_i\}$, the state and measurement operators can be parametrized as
\begin{equation}
    \label{parameterize}
    \rho=\sum_{i=1}^{d^2}\Theta_iE_i, \quad P_i=\sum_{j=1}^{d^2}\phi_{i,j} E_j,
\end{equation}
where $\Theta_i=\operatorname{Tr}(\rho E_i)$, $\phi_{i,j}=\operatorname{Tr}(P_iE_j)$. We then define $\Theta = [\Theta_1,\cdots,\Theta_{d^2}]^{T}$ and $\Phi_i=[\Phi_{i,1}, \cdots, \Phi_{i,d^2}]^{T}$.

We consider $J$ distinct POVM elements $\{P_i\}_{i=1}^{J}$, grouped into $M$ measurement configurations, each repeated $N/M$ times so that the total number of copies is $N$.
This allows us to formulate the QST task as a linear regression problem~\citep{Qi_2013}. 
The measurement model takes the form
\begin{equation}
    Y = X \Theta + e,
\end{equation}
where
\begin{equation}
    Y = [p_1, \cdots, p_J]^{T}, 
    X = [\Phi_1, \cdots, \Phi_J]^{T}, 
    e = [e_1, \cdots, e_J]^{T}.
\end{equation}
Here, $X$ is the parametrization matrix associated with the measurement 
operators $\{P_i\}_{i=1}^{J}$, and $Y$ and $e$ denote the measurement 
outcome vector and the corresponding error vector, respectively.
Assuming that the measurement operators are informationally complete, the linear equation has the standard least squares solution
\begin{equation}
    \label{lre_ls}
    \hat{\Theta}_{LS}=(X^{T} X)^{-1}X^{T} Y.
\end{equation}

Using the parameters in $\hat{\Theta}_{LS}$, we reconstruct the estimated state $\tilde{\rho}$. The upper bound on the MSE of $\hat{\Theta}_{LS}$~\citep{Qi_2013} is
\begin{equation}
    \label{lre_mse}
    \mathbb{E}\left[(\hat{\Theta}_{LS}-\Theta)^{T}(\hat{\Theta}_{LS}-\Theta)\right] \leq \frac{M}{4N} \operatorname{Tr}\left[ (X^{T} X)^{-1} \right].
\end{equation}

Note that the current state estimate may not be positive semidefinite, and can be corrected to a physical estimate with algorithms, such as one presented by \cite{Smolin_2012}.

\subsection{Dynamic tomography via observable evolutions}\label{qst_ott}

Building on the static LRE formulation in Section~\ref{qst_lre}, we consider a dynamic QST approach based on state evolution. In this setting, \cite{Xiao_2024} uses observable time traces to identify the initial quantum state of a closed quantum system. We start from the time-varying state $\rho(t)$ and its evolution dynamics described by the Liouville-von Neumann equation~\citep{Nielsen_Chuang_2010}
\begin{equation}
    \label{LvN}
    \dot{\rho}(t)=\text{i}[H,\rho(t)],
\end{equation}
where $[A,B]=AB-BA$ is the commutator, and $H$ is the traceless Hamiltonian. We define the structure constants $\Omega_{jlk}\in\mathbb{R}$ of the Lie algebra $\mathfrak{su}(d)$ \citep{Zhang_2014} with respects to the orthonormal basis $\{\text{i}E_j\}_{j=1}^{d^2-1}$ as
\begin{equation}
    \label{struct_const}
    [\text{i}E_j,\text{i}E_l]=\sum_{k=1}^{d^2-1}\Omega_{jl k}\,\text{i}E_k.
\end{equation}

To measure the state, here we use one traceless observable $O$, yielding measurement results $y(t)=\operatorname{Tr}(O\rho(t))$. Using the chosen basis, the state, Hamiltonian, and observable can be parametrized as
\begin{equation}
    \label{dynamic_para}
    \begin{gathered}
        \rho(t)= \frac{1}{\sqrt{d}}E_{d^2}+ \sum_{k=1}^{d^2-1}x_k(t)E_k, \\ \quad H=\sum_{k=1}^{d^2-1}h_kE_k, \quad O=\sum_{k=1}^{d^2-1}C_kE_k.
    \end{gathered}
\end{equation}
We denote $x=[x_1, \cdots, x_{d^2-1}]^{T}$, often referred to as the coherence vector~\citep{alicki,Zhang_2014}, $h=[h_1, \cdots, h_{d^2-1}]^{T}$, and $C=[C_1 ,\cdots, C_{d^2-1}]$.

Using~\eqref{LvN}, we define the matrix $A_c$ \citep{alicki,Zhang_2014}, such that
\begin{equation}
    \label{A_c}
    (A_c)_{jk}=\sum_{m=1}^{d^2-1}\Omega_{mjk}\,h_m,\quad A_c^{T}=-A_c.
\end{equation}
Let the system have an initial state $\rho_0$, parametrized as $x_0$. As such, the continuous system is
\begin{equation}
    \label{ott_cont}
    \begin{cases}
        \dot{x}(t)=A_cx(t),\\
        y(t) = Cx(t),\\
        x(0)=x_0.
    \end{cases}
\end{equation}

When sampled at discrete time intervals, the system can be discretized and described as
\begin{equation}
    \label{ott_sys}
    \begin{cases}
        x(k+1)=Ax(k),\\
        y(k) = Cx(k),\\
        x(0)=x_0,
    \end{cases}
\end{equation}
where $k$ denotes the current time step, sampled at a set interval $\Delta t$, with a total of $s$ samples, and $A=\exp(A_c \Delta t)$ is an orthogonal matrix \citep{Zhang_2014}. The system can therefore be expressed in a linear equation
\begin{equation}
    \label{ott_no_e}
    y=\mathcal{O}x_0,
\end{equation}
where $y=[y(0), \cdots, y(s-1)]^{T}$, and
\begin{equation}\label{obs1}
    \mathcal{O}=\left[\begin{array}{c}
C \\
C A \\
\vdots \\
C A^{n-1}
\end{array}\right]
\end{equation}
 captures system evolution in what is often considered the observability matrix in classical control theory~\citep{Callier_1991}. Considering experimental noise, denoting noisy measurement results as $\hat{y}$ with measurement error vector $e$, we have the unified linear equation
\begin{equation}
    \label{ott}
    \hat{y}=\mathcal{O}x_0+e.
\end{equation}
If $x_0$ is identifiable, which is true if and only if $\operatorname{rank}(\mathcal{O})=d^2-1$, the system has standard least square solution
\begin{equation}
    \label{ott_ls}
    \hat{x}_0 = (\mathcal{O}^{T} \mathcal{O})^{-1}\mathcal{O}^{T}\hat{y}.
\end{equation}
Since we only use one observable, similar to the base LRE formulation~\eqref{lre_ls}, the upper bound on the MSE of $\hat{x}_0$ is 
\begin{equation}
    \label{ott_mse}
    \mathbb{E}\left[(\hat{x}_{0}-x_0)^{T}(\hat{x}_{0}-x_0)\right] \leq \frac{s}{4N} \operatorname{Tr}\left[ (\mathcal{O}^{T} \mathcal{O})^{-1} \right].
\end{equation}

\section{Robustness Analysis of Linear Regression Estimation}\label{sec3}
In this section we describe and analyse the QST robustness problem addressed by this paper. We denote perturbation errors using $\Delta$ such that $\Delta(\cdot) = \hat{(\cdot)}  - (\cdot)$, where $\hat{}$ indicates a perturbed item.

\subsection{Error bound for static tomography}\label{pa_ea_lre}
Given the QST algorithm described in Section \ref{qst_lre}, we examine robustness under POVM set perturbations, quantified by shifts in the state estimate MSE upper bound. Assume that there is some perturbation in the measurement operators $P_i$, such that 
\begin{equation}
    \label{static_op_error}
    ||\hat{P}_i-P_i||_F = ||\Delta P_i||_F\leq \epsilon_P,
\end{equation}
where $||\cdot||_F$ is the Frobenius norm.

 We then propose the following theorem to characterize the MSE bound with the measurement perturbations: 
\begin{theorem}
 If  $\lambda_{d^2} - 2\sqrt{J}\epsilon_P||X||_F\geq 0$ where $\lambda_{d^2}$ is the minimal eigenvalues of $X^{T}X$, we have
\begin{equation}
    \begin{aligned}  
        &\left| \frac{M}{4N}\operatorname{Tr}[(\hat{X}^{T} \hat{X})^{-1}]-\frac{M}{4N}\operatorname{Tr}[(X^{T} X)^{-1}] \right| \\ 
        \leq &
        \frac{M\sqrt{J}d^2 \epsilon_P||X||_F}{2N\lambda_{d^2} (\lambda_{d^2} - 2\sqrt{J}\epsilon_P||X||_F)}
   \end{aligned}
\end{equation}
\end{theorem}

\textbf{Proof.} To determine the error bound of the perturbed system, we start with the perturbed measurement operators $P_i$ and derive the original MSE bound expression~\eqref{lre_mse}.

First, we consider the error in $P_i$. Using~\eqref{parameterize}, ~\eqref{static_op_error}, and the Cauchy-Schwarz inequality we obtain that
\begin{equation}
    \label{Delta_Phi_bound}
    |\Delta \Phi^i_k|\leq ||\Delta P_i||_F ||E_k||_F \leq \epsilon_P.
\end{equation}

Secondly, we consider the error in $X$, 
\begin{equation}
    \label{Delta_X_norm}
    \begin{aligned}
        ||\Delta X||_F^2 &= \sum_{i=1}^{J}||\hat{\Phi}_i-\Phi_i||_F^2 \leq J\epsilon_P^2.  
    \end{aligned}
\end{equation}
Next, we consider the error in $X^{T}X$,
\begin{equation}
    \label{Delta_XTX_norm}
    \begin{aligned}
        ||\Delta ({X^{T} X})||_F 
        &= ||X^{T} \Delta X + \Delta X^{T} X+\Delta X^{T} \Delta X||_F\\
        &\leq ||X^{T}\Delta X||_F + ||\Delta X^{T} X||_F + ||\Delta X^{T} \Delta X||_F\\
        &\leq 2||X||_F \cdot ||\Delta X||_F + ||\Delta X||_F^2\\
        &\leq 2\sqrt{J}\epsilon_P||X||_F+J\epsilon_P^2\sim 2\sqrt{J}\epsilon_P||X||_F.\\
    \end{aligned}
\end{equation}
Now we consider the trace expression $\operatorname{Tr}[(X^{T} X)^{-1}]$. To begin, we introduce \textit{Weyl's} perturbation theorem~\citep{Bhatia_1997} in the following Lemma:
\begin{lemma}
    \label{lemma_weyl}
    Let $A$, $B$ be Hermitian matrices with eigenvalues sorted in descending order, such that $\lambda_1(A) \geq \cdots \geq \lambda_n(A)$ and $\lambda_1(B) \geq \cdots \geq \lambda_n(B)$, respectively. Then,
    \begin{equation}
        \label{lemma_weyl_eq}
        \max_j|\lambda_j(A)-\lambda_j(B)|\leq ||A-B||_2\leq ||A-B||_F.
    \end{equation}
\end{lemma}
The Euclidean norm for a vector and the spectral norm for a matrix are defined as $||\cdot||_2$.
Let $X^{T} X$ have eigenvalues $\lambda_1 \geq \cdots \geq\lambda_{d^2} >0$ and let $\hat{X}^{T} \hat{X}$ have eigenvalues $\mu_1 \geq\cdots\geq\mu_{d^2}$, such that
\begin{equation}
    \label{eq:XTX_weyl}
    |\mu_i-\lambda_i| \leq ||\hat{X}^{T} \hat{X} - X^{T} X||_F,\quad i=1,\cdots,d^2.
\end{equation}
If  $\lambda_{d^2} - 2\sqrt{J}\epsilon_P||X||_F\geq 0$, we have $\mu_i \geq \lambda_{d^2} - 2\sqrt{J}\epsilon_P||X||_F\geq 0$ for all $i$ and thus $\hat{X}^{T} \hat{X}$ is always a positive definite matrix.

Using \eqref{Delta_XTX_norm} and \eqref{eq:XTX_weyl}, $|\operatorname{Tr}[(X^{T} X)^{-1}]|$ can be bounded as
\begin{equation}\label{eq:Delta_Tr_block}
    \begin{aligned}
        &|\Delta\operatorname{Tr}[(X^{T} X)^{-1}]|
        = \left| \sum_{i=1}^{d^2}\left( 
        \frac{1}{\mu_i}-\frac{1}{\lambda_i} \right) \right|\\
        \leq &\left( 
        \frac{\sum_{i=1}^{d^2} \left| \lambda_i - \mu_i \right|}{\lambda_{d^2} \mu_{d^2}} \right)
        \leq \left( 
        \frac{d^2 ||\Delta ({X^{T} X})||_F}{\lambda_{d^2} \mu_{d^2}} \right)\\
        \leq& \left( 
        \frac{2\sqrt{J} d^2 \epsilon_P||X||_F}{\lambda_{d^2} \mu_{d^2}} \right).
    \end{aligned}
\end{equation}

Using \eqref{eq:XTX_weyl} and \eqref{eq:Delta_Tr_block}, we obtain the final expression
\begin{equation}
\label{eq:Delta_Tr_bound}
    |\Delta\operatorname{Tr}[(X^{T} X)^{-1}]| 
    \leq \left( \frac{2\sqrt{J}d^2 \epsilon_P||X||_F}{\lambda_{d^2} (\lambda_{d^2} -  2\sqrt{J}\epsilon_P||X||_F)} \right).
\end{equation}

Ultimately, we obtain the following upper bound that describes the shift in the MSE upper bound due to the influence of measurement operator perturbations,
\begin{equation} \label{lre_robust}
    \begin{aligned}  
        &\left| \frac{M}{4N}\operatorname{Tr}[(\hat{X}^{T} \hat{X})^{-1}]-\frac{M}{4N}\operatorname{Tr}[(X^{T} X)^{-1}] \right| \\ 
        \leq &    
        \frac{M\sqrt{J}d^2 \epsilon_P||X||_F}{2N\lambda_{d^2} (\lambda_{d^2} - 2\sqrt{J}\epsilon_P||X||_F)}.
   \end{aligned}
\end{equation} \hfill $\Box$

The upper bound in \eqref{lre_robust} depends on the perturbation magnitude of 
the measurement operators $\epsilon_P$, the ideal measurement operators, and 
the system dimension $d$. We note that the bound is generally conservative and 
not tight, but it characterizes the worst-case scenario.

\subsection{Error bound for dynamic tomography}\label{pa_ea_ott}
Given the QST algorithm described in Section~\ref{qst_ott}, we now examine its robustness under perturbations of the observable $O$ and the Hamiltonian $H$. The perturbations are first considered separately, and then jointly. We again quantify the error using the MSE upper bound shift.


In the observable $O$, assume that there is some perturbation such that 
\begin{equation}
    \label{dynamic_O_C_error}
    ||\hat{C}-C||_2=||\hat{O}-O||_F \leq \epsilon_O.
\end{equation}
Let the traceless assumption for the observable also hold after perturbation, such that $\operatorname{Tr}(\hat{O})=0$.
In the Hamiltonian $H$, assume that there is some perturbation such that
\begin{equation}
    \label{dynamic_H_error}
    ||\hat{h}-h||_2 = ||\hat{H}-H||_F \leq \epsilon_H.
\end{equation}

We then present the following theorem to characterize the MSE bound with the observable and Hamiltonian perturbations:
\begin{theorem}
 If  $\lambda_{d^2-1} - 2   ||\Delta \mathcal{O}||_F||\mathcal{O}||_F\geq 0$ where $\lambda_{d^2-1}$ is the minimal eigenvalues of $\mathcal{O}^{T}\mathcal{O}$ and $||\Delta \mathcal{O}||_F$ is defined in \eqref{DOb_both}, we have
\begin{equation}
\begin{aligned}
        &\left| \frac{s}{4N}\operatorname{Tr}\left[ (\hat{\mathcal{O}}^{T} \hat{\mathcal{O}})^{-1} \right] - \frac{s}{4N}\operatorname{Tr}\left[ (\mathcal{O}^{T} \mathcal{O})^{-1} \right] \right| \\
        \leq& 
         \frac{s(d^2-1)||\mathcal{O}||_F||\Delta\mathcal{O}||_F}{2N\lambda_{d^2-1} (\lambda_{d^2-1} - 2\,||\mathcal{O}||_F\,||\Delta\mathcal{O}||_F)}.
\end{aligned}
\end{equation}
\end{theorem}

\textbf{Proof.} To determine the error bound of the perturbed system, we start with each perturbation separately and then combine the effects when both perturbations are active.

First we consider perturbations in $O$, which leads to the perturbed observability matrix $\hat{\mathcal{O}}=[\hat{C} \ \hat{C}A \ \cdots \ \hat{C}A^{s-1}]^{T}$ like \eqref{obs1}. Since $A$ is orthogonal due to the anti-symmetry of $A_c$, we have $||A^k||_2=1$, for $k=0,\cdots,s-1$, and therefore
\begin{equation}
    \label{ott_C_O}
    \begin{aligned}
        ||\Delta \mathcal{O} ||^2_F
        &=\sum_{k=0}^{s-1}||\Delta (CA^k)||_2^2 =s||\Delta C||^2_2  \leq s \epsilon_O^2.
    \end{aligned}
\end{equation}
As such, it holds that $||\Delta \mathcal{O}||_F\leq \sqrt{s}\epsilon_O$. Structurally, this is much like its static LRE counterpart~\eqref{Delta_X_norm}.

Next we consider perturbations in $H$, which leads to the perturbed observability matrix $\hat{\mathcal{O}}=[C \ C\hat{A} \ \cdots \ C\hat{A}^{s-1}]^{T}$ like \eqref{obs1}. We begin by relating $\Delta H$ to $\Delta A$. Using \eqref{dynamic_para}, \eqref{A_c}, and $\sum_{j,k} \Omega_{mjk} \Omega_{njk} = 2d \delta_{mn}$~\citep{Hall2015Lie} where $\delta$ is Kronecker delta function, we have
\begin{equation}
    \label{DAc}
    \begin{aligned}
        ||\Delta A_c||_F^2
        &= \sum_{j,k}\Big(\sum_m \Omega_{mjk}(\Delta h)_m\Big)^2\\
        &= \sum_{j,k}\Big(\sum_m \Omega_{mjk}(\Delta h)_m\Big)  \Big(\sum_n \Omega_{njk}(\Delta h)_n\Big) \\
        &= \sum_{m,n}\Big( \sum_{j,k}\Omega_{mjk}\Omega_{njk} \Big)( \Delta h)_m(\Delta h)_n\\
        &\leq 2d\epsilon_H^2.
    \end{aligned}
\end{equation}
Now we consider the error in $A$ using an interpolating product for Duhamel's formula~\citep{Pazy1983Semigroups}, as 
\begin{equation}
    \label{g_tau}
    g(\tau)=e^{(1-\tau)A_t}e^{\tau(A_t+\Delta A_t)},
\end{equation}

where $A_t = A_c \Delta t$, $\Delta A_t = \Delta A_c \Delta t$, $\tau\in [0,1]$, such that $g(0)=e^{A_t}=A$, $g(1)=e^{A_t+\Delta A_t}=\hat{A}$. Taking the derivative yields $\frac{d}{d\tau}g(\tau)=e^{(1-\tau)A_t}\Delta A_t e^{\tau(A_t+\Delta A_t)}$, and thus we can evaluate the integral as
\begin{equation}
    \begin{aligned}
        \hat{A}-A
        &=\int_0^1 e^{(1-\tau)A_t}\Delta A_te^{\tau(A_t+\Delta A_t)} d\tau,
    \end{aligned}
\end{equation}
Therefore,
\begin{equation}
    \label{DA}
    \begin{aligned}
        ||\hat{A}-A||_2
        &\leq \int_0^1 ||e^{(1-\tau)A_t}||_2 \; ||\Delta A_t||_2 \; ||e^{\tau(A_t+\Delta A_t)}||_2\; d\tau\\
        &= \int_0^1 ||\Delta A_t||_2 \ d\tau=||\Delta A_t||_2=\Delta t||\Delta A_c||_2\\
        &\leq \sqrt{2d} \Delta t  \epsilon_H.
    \end{aligned}
\end{equation}

We then return to the observability matrix. The Fr\'echet derivative~\citep{Higham2008Functions, AlMohy2009FrechetExp} of a power gives, for $k \geq 1$, 
\begin{equation}
    \label{frechet}
    \begin{aligned}
        \Delta(CA^k)
        &=C\sum_{j=0}^{k-1}A^{k-1-j}\Delta A A^j.\\
    \end{aligned}
\end{equation}
It follows that $||\Delta(CA^k)||_2 \leq k||C||_2||\Delta A||_2$. We then have
\begin{equation}
    \label{DO}
    ||\Delta\mathcal{O}||^2_F\leq \sum_{k=0}^ {s-1} ||\Delta(CA^k)||_2^2 \leq ||C||_2^2 \; || \Delta A||_2^2 \left( \sum_{k=1}^{s-1}k^2 \right).
\end{equation}

Using \eqref{DA}, \eqref{DO} and $\sum_{k=1}^{s-1}k^2=\frac{(s-1)s(2s-1)}{6}$, we obtain 
\begin{equation}
    \label{DO_fin}
    \begin{aligned} 
    ||\Delta\mathcal{O}||_F&\leq ||C||_2 \Delta t  \left(\frac{2d(s-1)s(2s-1)}{6}\right)^{\frac{1}{2}} \epsilon_H.
     \end{aligned}
\end{equation}

We now consider the case when both perturbations are active. Starting from the observability matrix block rows, using \eqref{DA}, \eqref{frechet} \eqref{dynamic_O_C_error}, and the product expansion $$\hat C\,\hat A^{\,k}-CA^{\,k} =\Delta C\,A^{\,k}+C(\hat A^{\,k}-A^{\,k})+\Delta C(\hat A^{\,k}-A^{\,k}),$$ 
we have
\begin{equation}
    \label{DOb_both}
  \begin{aligned}
        &||\Delta\mathcal{O}||^2_F=
        \sum_{k=0}^{s-1}||\hat C\,\hat A^{\,k}-CA^{\,k}||_2^2 \\
        \leq& 
        s\,||\Delta C||^2_2
        + \left(\sum_{k=1}^{s-1}k^2\right)(||C||_2+||\Delta C||_2)^2\,||\Delta A||^2_2\\
        \leq&
        s\epsilon_O^2+\left( \frac{(s-1)s(2s-1)}{6} \right)(||C||_2+\epsilon_O)^2 (\Delta t)^22d\epsilon_H^2\\
        \sim&  s\epsilon_O^2+2d s^{3}||C||_{2}^2 (\Delta t)^2\epsilon_H^2.
 \end{aligned}
\end{equation}
If  $\lambda_{d^2-1} - 2   ||\Delta \mathcal{O}||_F||\mathcal{O}||_F\geq 0$, we can also ensure that $\hat{\mathcal{O}}^{T}\hat{\mathcal{O}}$ is positive definite. Therefore, we can now again apply similar procedures to Section \ref{pa_ea_lre} to have the following relationship,
\begin{equation}
    \label{rob_bound}
\begin{aligned}
        &\left| \frac{s}{4N}\operatorname{Tr}\left[ (\hat{\mathcal{O}}^{T} \hat{\mathcal{O}})^{-1} \right] - \frac{s}{4N}\operatorname{Tr}\left[ (\mathcal{O}^{T} \mathcal{O})^{-1} \right] \right| \\
        \leq& 
         \frac{s(d^2-1)||\mathcal{O}||_F||\Delta\mathcal{O}||_F}{2N\lambda_{d^2-1} (\lambda_{d^2-1} - 2\,||\mathcal{O}||_F\,||\Delta\mathcal{O}||_F)},
\end{aligned}
\end{equation}
where $||\Delta\mathcal{O}||_F$ is given by \eqref{DOb_both}.

The upper bound in \eqref{rob_bound} depends on the perturbation magnitude of the
observable and Hamiltonian, the ideal observable and Hamiltonian, the sampling time $\Delta t$, and 
the system dimension $d$. Similarly, the bound is not tight and characterizes the worst-case scenario.

\section{Numerical Validation}\label{sec4}

\subsection{Static tomography simulation}\label{nv_sts}
We begin by constructing a random mixed unknown state $\rho$. For the single qubit case, we use the Pauli matrices~\citep{Nielsen_Chuang_2010} and the two-dimensional identity matrix,
\begin{equation}
    \label{paulis}
    \sigma_x=\begin{bmatrix}
        0 & 1 \\ 1 & 0
    \end{bmatrix},
    \;
    \sigma_y=\begin{bmatrix}
        0 & -i \\ i & 0
    \end{bmatrix},
    \;
    \sigma_z=\begin{bmatrix}
        1 & 0 \\ 0 & -1
    \end{bmatrix},
    \;
    I_2=\begin{bmatrix}
        1 & 0 \\ 0 & 1
    \end{bmatrix}.
\end{equation}

For measurement, we implement the Cube measurement set~\citep{PhysRevA.78.052122}, defined for a single qubit system as $\{\frac{I\pm\sigma_x}{2}, \frac{I\pm\sigma_y}{2}, \frac{I\pm\sigma_z}{2}\}$. For multi-qubit systems, the Cube measurements are the tensor products of one-qubit Cube measurements. Random perturbations are applied to each POVM element, while still enforcing POVM constraints and theorem conditions on $\epsilon_P$. Results are averaged over $100$ trials per setting.

\begin{figure}[h]
    \centering
    \includegraphics[width=1\linewidth]{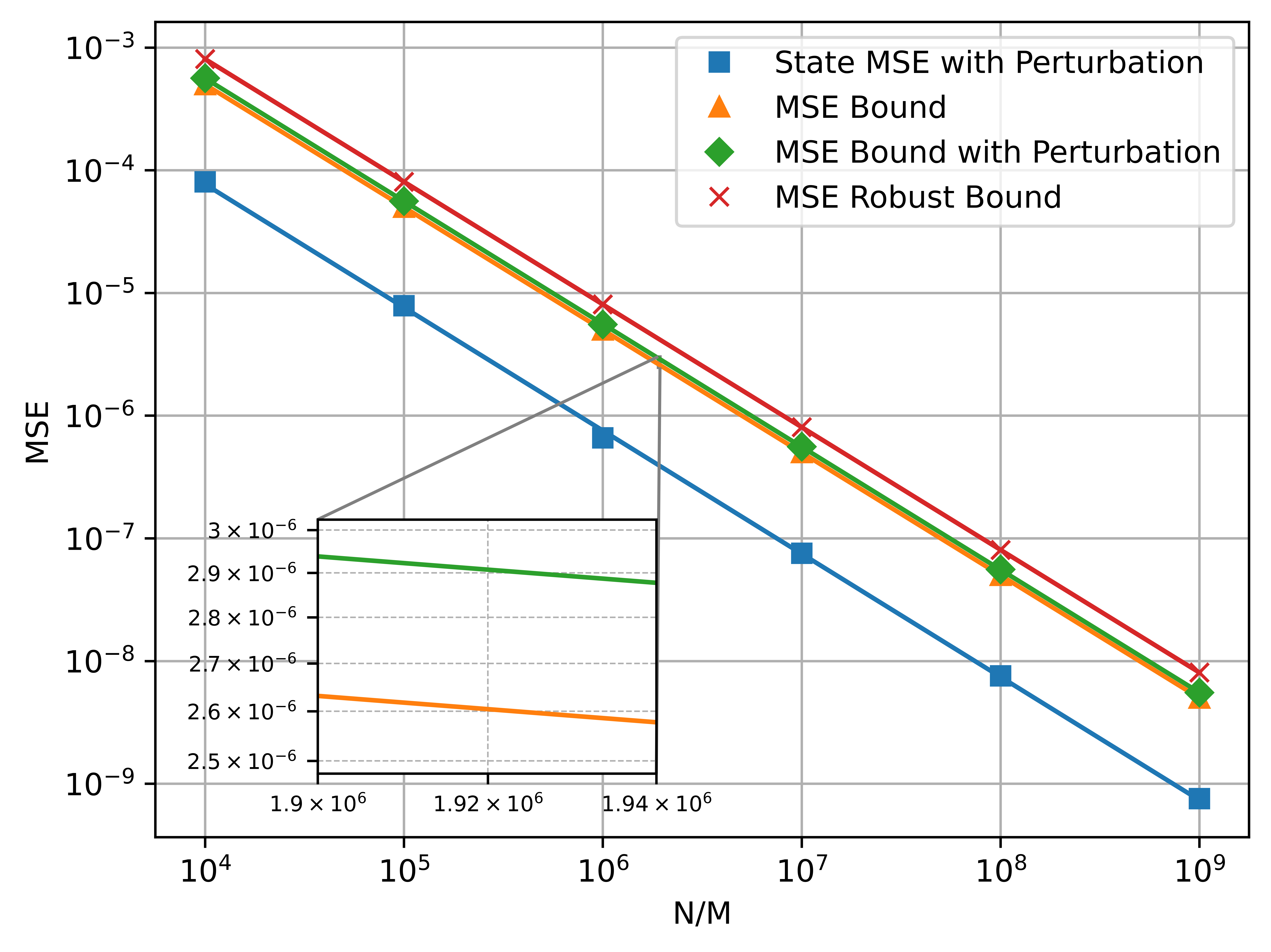}
    \caption{Single-qubit static tomography including the state estimation MSE under perturbed POVMs and the corresponding nominal, perturbed, and robust MSE upper bounds vs resources per detector $N/M$ in~\eqref{lre_robust}.}
    \label{fig:static_N}
\end{figure}

\begin{figure}[h]
    \centering
    \includegraphics[width=1\linewidth]{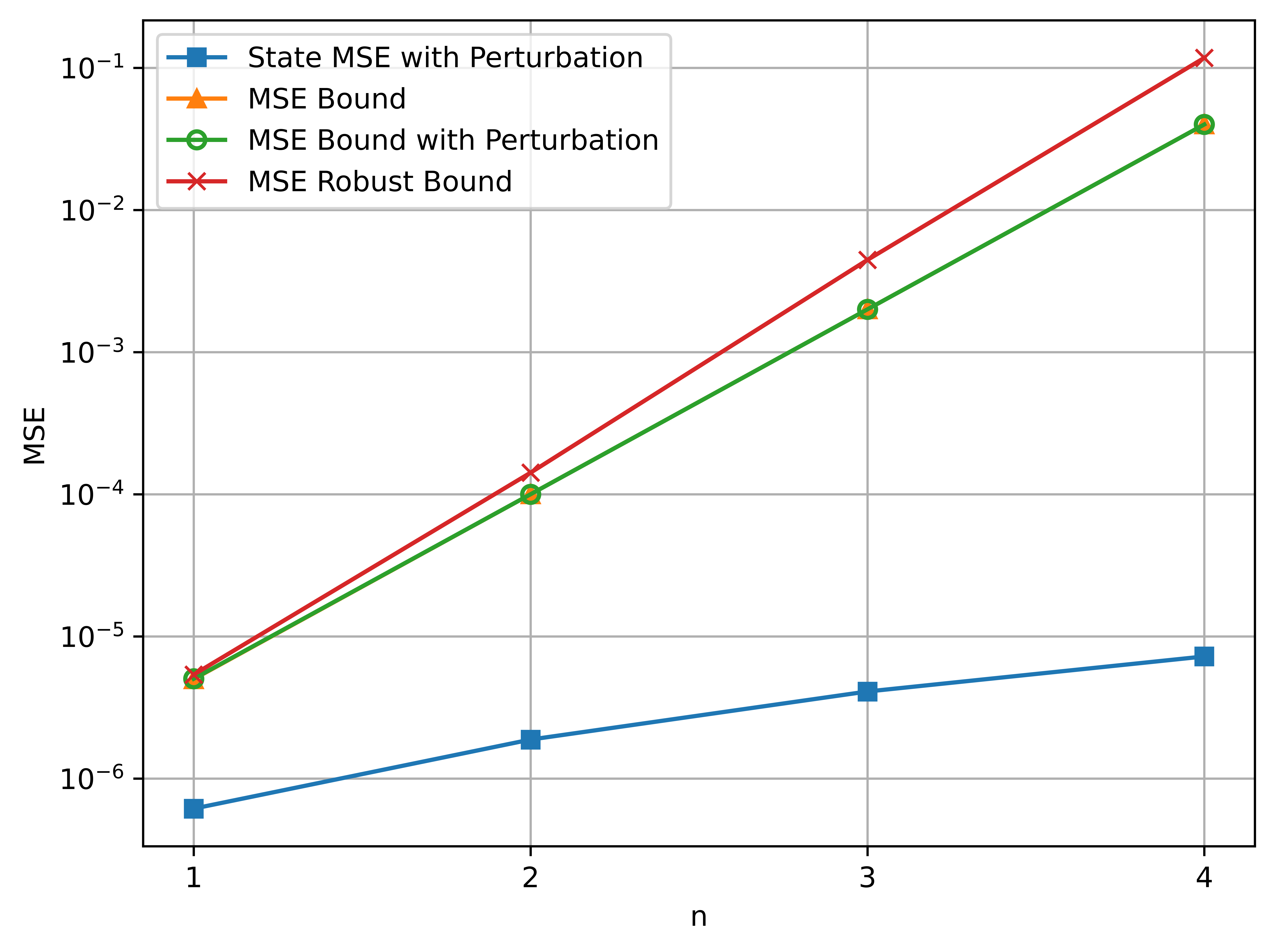}
    \caption{Static tomography for an $n$-qubit system at fixed resources per detector $N/M=10^6$, including state estimation MSE and associated nominal, perturbed, and robust MSE bounds as a function of the number of qubits.}
    \label{fig:static_n}
\end{figure}

Data and legends in Fig.~\ref{fig:static_N} and Fig.~\ref{fig:static_n} include the following:
\begin{itemize}
    \item State MSE with Perturbation: True MSE with perturbed measurements.
    \item MSE Bound: $\frac{M}{4N}\operatorname{Tr}[(X^{T} X)^{-1}]$.
    \item MSE Bound with Perturbation: $\frac{M}{4N}\operatorname{Tr}[(\hat{X}^{T} \hat{X})^{-1}]$.
    \item MSE Robust Bound: Original MSE bound extended as described by~\eqref{lre_robust}, with equality.
\end{itemize}

In Fig.~\ref{fig:static_N}, we consider single-qubit tomography with $M=3$. 
The state estimation MSE exhibits the expected $O(1/N)$ scaling with respect 
to the total resource $N$, and the robust bound consistently remains above 
both the nominal and perturbed bounds. 
In Fig.~\ref{fig:static_n}, we plot the MSE as a function of the number of 
qubits $n$, with $M = 3^{n}$. Both the state estimation MSE and the 
corresponding bounds increase with $n$, reflecting the growth in system 
dimension and the larger number of parameters to be estimated. The robust 
bound also becomes increasingly conservative as the dimension increases.

\subsection{Dynamic tomography simulation}
\begin{figure}[h]
    \centering
    \includegraphics[width=1\linewidth]{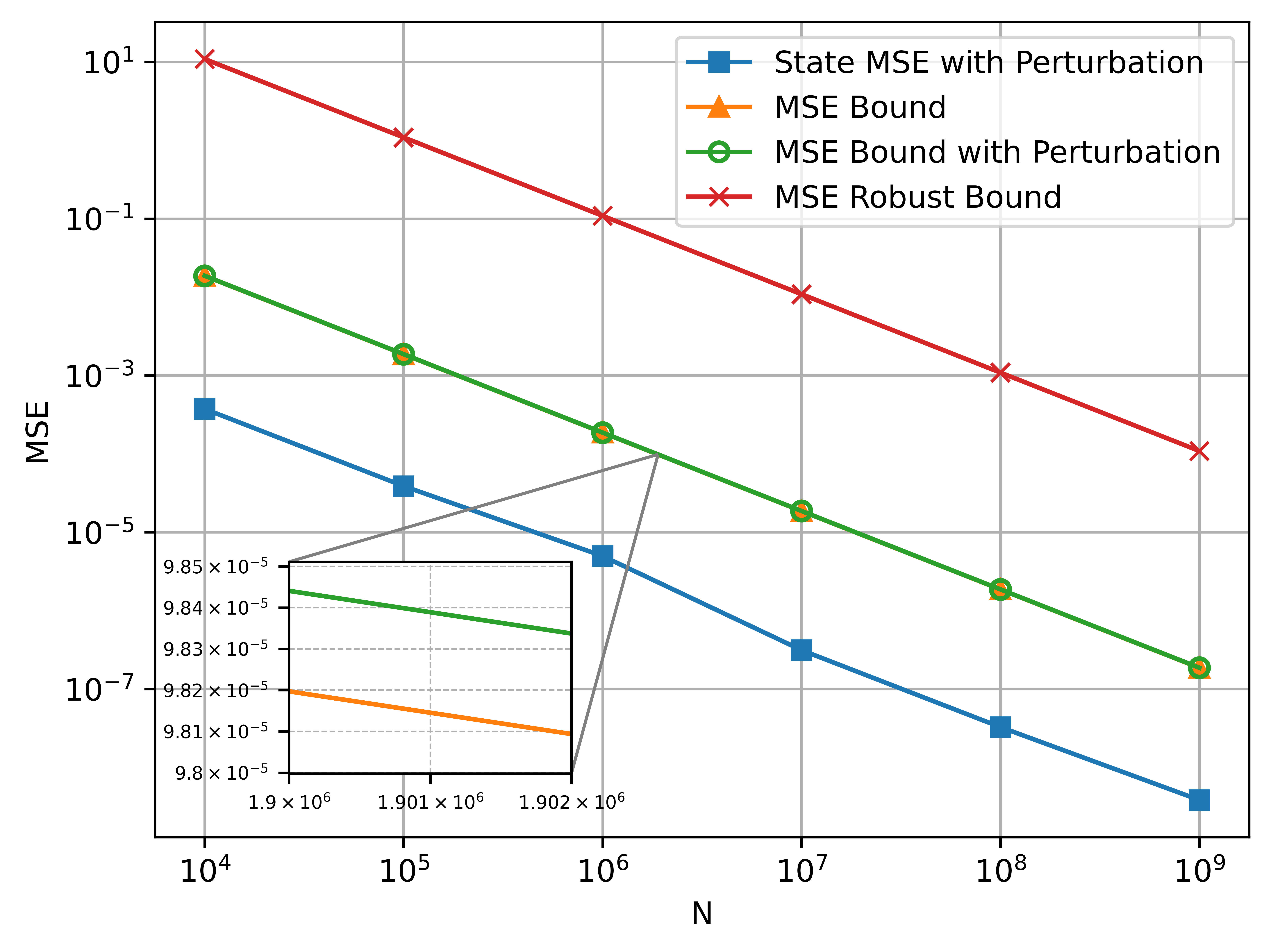}
    \caption{Dynamic tomography for a single qubit, including the state estimation MSE under simultaneous perturbations of both the observable and the Hamiltonian, along with the nominal, perturbed, and robust MSE bounds in~\eqref{rob_bound}.}
    \label{fig:dynamic}
\end{figure}

For the dynamic case, we consider a single qubit system with observable $O=\sigma_x$ and Hamiltonian $H = 0.2\sigma_x + 0.3\sigma_y + 0.5\sigma_z$, sampling at $\Delta t=0.05$ for a total of $s=24$ samples. Data in Fig.~\ref{fig:dynamic} includes the same legend conventions as the static simulation in Section~\ref{nv_sts}, except that the MSE Robust Bound is now computed using~\eqref{rob_bound}.

In Fig.~\ref{fig:dynamic}, the state estimation MSE again scales at $O(1/N)$, and all bounds remain below the robust bound. The inset highlights the small difference between the nominal and perturbed bounds. This is consistent with the eigenvalue condition on $\mathcal{O}^{T}\mathcal{O}$, which forces the admissible perturbation to be very small for a single observable. This again shows that the robust bound is conservative, and captures a worst-case perturbation under a sufficient but not necessary condition on $\epsilon_O$ and $\epsilon_H$.

\section{Conclusion}\label{sec5}

In this paper, we investigated the robustness of quantum state tomography in static and dynamic settings within the LRE framework. For static tomography, we analyzed how bounded perturbations in the measurement operators affect the standard MSE bound and derived an explicit robust counterpart. For dynamic tomography, we carried out an analysis for simultaneous perturbations in the observable and Hamiltonian. Numerical simulations for qubit systems showed how these bounds scale with the number of copies and system dimension, and confirmed that the empirical MSE remains below the robust bounds. Future work includes tightening these bounds and extending the dynamic analysis to multiple observables and higher-dimensional systems.

\bibliography{rank}             

\end{document}